\documentclass[aps,prl,twocolumn]{revtex4}
\usepackage{graphicx} 
\usepackage{amsmath}
\usepackage{color}
\newcommand{\be}{\begin{equation}}
\newcommand{\ee}{\end{equation}}
\newcommand{\bea}{\begin{eqnarray}}
\newcommand{\eea}{\end{eqnarray}}
\newcommand{\bse}{\begin{subequations}}
\newcommand{\ese}{\end{subequations}}
\newcommand{\ecp}{${\rm EuCo_2P_2}$}
\newcommand{\bcp}{${\rm BaCo_2P_2}$}

\begin{document}

\title{${\rm\bf EuCo_2P_2}$: A Model Molecular-Field Helical Heisenberg Antiferromagnet}

\author{N. S. Sangeetha}
\author{Abhishek Pandey}
\altaffiliation{Present address: Department of Physics and Astronomy, Louisiana State University, Baton Rouge, Louisiana 70803, USA}
\author{D. C. Johnston}
\email{johnston@ameslab.gov}
\affiliation{Ames Laboratory and Department of Physics and Astronomy, Iowa State University, Ames, Iowa 50011, USA}

\date{\today}

\begin{abstract}

The Eu spins-7/2 in \ecp\ with the tetragonal ${\rm ThCr_2Si_2}$-type structure have the prototypical noncollinear helical antiferromagnetic structure below the N\'eel temperature $T_{\rm N} = 66.5$~K as previously determined from single-crystal neutron diffraction measurements.  The helix axis is along the $c$~axis with the ordered moments aligned within the $ab$~plane.  Our recent formulation of molecular field theory (MFT) is found to quantitatively fit the anisotropic magnetic susceptibility of single-crystal \ecp\ at $T\leq T_{\rm N}$ with a helix turn angle comparable to the neutron diffraction value.  The experimental and MFT magnetic heat capacities at $T\leq T_{\rm N}$ are also in good agreement.  Values of the Heisenberg exchange interactions between the Eu spins-7/2 within the $J_0$-$J_{z1}$-$J_{z2}$ MFT model are derived.  The results demonstrate the robust applicability of the MFT to model the thermodynamic properties of noncollinear Heisenberg antiferromagnets.

\end{abstract}


\maketitle


We recently presented a general formulation of the Weiss molecular field theory (MFT) of local-moment Heisenberg antiferromagnets (AFMs) containing identical crystallographically-equivalent spins \cite{Johnston2012, Johnston2015, Johnston2015b}.  In this formulation the paradigm of magnetic sublattices previously required to calculate the magnetic and thermal properties below the AFM ordering (N\'eel) temperature $T_{\rm N}$ was abandoned and thermodynamic predictions for various systems were obtained from consideration of only the exchange interactions between a central moment and its neighbors.  This formulation allows the anisotropic temperature~$T$-dependent magnetic susceptibility $\chi$ and the magnetic heat capacity $C_{\rm mag}(T)$ of collinear and planar noncollinear AFMs at $T\leq T_{\rm N}$ to be calculated on the same footing.  Furthermore, the theoretical predictions are cast in terms of measurable properties.  For collinear AFMs these are the spin~$S$, $T_{\rm N}$, $\chi(T_{\rm N})$ and the Weiss temperature $\theta_{\rm p}$ in the Curie-Weiss law for $\chi$ in the paramagnetic state.  For planar noncollinear AFMs an additional parameter is required.  In the $J_0$-$J_{z1}$-$J_{z2}$ MFT model for a helical AFM structure \cite{Johnston2012, Johnston2015}, the additional parameter is the turn angle $kd$ between ordered moments in adjacent magnetic moment layers along the helix axis.  Several  applications of this MFT to fit experimental $\chi(T\leq T_{\rm N})$ data for single crystals and polycrystals of collinear and planar noncollinear AFM systems have been presented \cite{Johnston2012, Johnston2015b, Goetsch2013, Samal2014, Anand2014, Nath2014, Goetsch2014, Anand2015, Bednarchuk2015}.

However, the MFT predictions have not yet been compared with single-crystal $C_{\rm mag}(T)$ and $\chi(T)$ data for any system with a (coplanar) helix AFM structure, which is the prototype for a noncollinear AFM structure.  The reason is that no such data were available for a helical structure in which the assumptions of the MFT are fulfilled, namely high spin (small quantum fluctuations), dominant Heisenberg spin interactions, and no ferromagnetic (FM) component or magnetic and/or spin reorientation transitions occurring below the initial $T_{\rm N}$.  Furthermore, for a definitive test of the applicability of the MFT predictions for a helix, one also requires input about the AFM structure from neutron diffraction measurements.

Single-crystal neutron diffraction measurements \cite{Reehuis1992} have shown that \ecp\ has the required helical AFM structure below $T_{\rm N} = 66.5$~K\@.  A generic picture of such a helix is shown in Fig.~1 of~\cite{Johnston2012}.  \ecp\ has the body-centered tetragonal (bct) ${\rm ThCr_2Si_2}$-type crystal structure (space group $I4/mmm$) \cite{Marchand1978}.  The lattice parameters of \ecp\ at room temperature are $a=3.7649(5)$~\AA\ and $c=11.348(2)$~\AA\ with $c/a=3.014(1)$ \cite{Marchand1978}.   The ordered moment at 15~K is $\langle\mu\rangle = 6.9(1)~\mu_{\rm B}$/Eu, which agrees with the saturation moment $\mu_{\rm sat} = gS\mu_{\rm B}$/Eu $= 7\,\mu_{\rm B}$/Eu expected for Eu$^{+2}$ spin $S=7/2$ and spectroscopic splitting factor $g=2$ \cite{Reehuis1992}.  The authors of \cite{Reehuis1992} reported that the AFM structure below $T_{\rm N}$ is a helix with the Eu ordered moments aligned in the $ab$~plane of the tetragonal structure, with the helix axis along the perpendicular $c$~axis.  The $ab$-plane alignment of the ordered moments is consistent with the prediction of magnetic dipole interactions between the Eu spins for the above $c/a$ ratio \cite{Johnston2016}.  The incommensurate AFM propagation vector for the helix changed by only 2\% from ${\bf k} = [0,\ 0,\ 0.834(4)]2\pi/c$ at $T=64$~K to $[0,\ 0,\ 0.852(4)]2\pi/c$ at $T=15$~K \cite{Reehuis1992}.  Since $d=c/2$ is the distance along the $c$~axis between adjacent layers of FM-aligned moments in the bct Eu sublattice, the turn angle $kd$ between the ordered moments in adjacent layers perpendicular to the helix axis is
\be
kd(64~{\rm K}) = 0.834(4)\pi\,{\rm rad},\quad  kd(15~{\rm K}) = 0.852(4)\pi\,{\rm rad}.
\label{Eq:kdNeuts}
\ee

Polycrystalline samples of \ecp\ exhibit AFM ordering of the Eu$^{+2}$ spins at $T_{\rm N} = 66.5(5)$~K as deduced from powder $\chi(T)$ data \cite{Morsen1988}.  The data in the paramagnetic state at $T>T_{\rm N}$ follow the Curie-Weiss law $\chi = \frac{C}{T-\theta_{\rm p}}$ where the positive $\theta_{\rm p}=20(2)$~K \cite{Morsen1988} indicates dominant FM interactions \cite{Johnston2012, Johnston2015}.  The effective moment obtained from the observed Curie constant is $\mu_{\rm eff} = 8.10(8)~\mu_{\rm B}$/f.u., where f.u.\ means formula unit.  This value is close to the value $\mu_{\rm eff}^{\rm calc} = g\sqrt{S(S+1)}\, \mu_{\rm B} =7.94~\mu_{\rm B}$ calculated from Eu$^{+2}$ spin $S=7/2$ with $g=2$. The authors' $^{151}$Eu M\"ossbauer data are also consistent with the oxidation state Eu$^{+2}$, and they inferred that Co does not contribute to the AFM ordering in \ecp\ \cite{Morsen1988}.  The values of $kd$ with $\pi/2 < kd < \pi$ in Eq.~(\ref{Eq:kdNeuts}) indicate that the dominant {\it interlayer} interactions are AFM [see Eq.~(\ref{Eq:coskd}) below], and the above $\theta_{\rm p}$ value therefore indicates that the dominant {\it intralayer} interactions must be FM [see Eq.~(\ref{eq:thetap}) below].  Electrical resistivity $\rho(T)$ measurements on single crystals indicate metallic character with a sharp break in slope at $T_{\rm N}$ \cite{Nakama2010}.

In order to test the applicability of our formulation of MFT to the noncollinear AFM state of \ecp\ we grew crystals of \ecp\ and measured their properties.  We find that this compound is a model system exhibiting an anisotropic $\chi(T)$ at $T\leq T_{\rm N}$ that is well described by the MFT for a helical AFM structure with turn angles comparable to the neutron diffraction results in Eq.~(\ref{Eq:kdNeuts}).  We also present $C_{\rm mag}(T)$ data and show that the MFT prediction is in good agreement with these data for $T\leq T_{\rm N}$\@.  An independent measurement of $T_{\rm N}$ was obtained from in-plane $\rho(T)$ measurements.  Finally, the Heisenberg exchange interactions between the Eu spins are estimated within a minimal $J_0$-$J_{z1}$-$J_{z2}$ MFT model.

\begin{figure}
\includegraphics[width=2.85in]{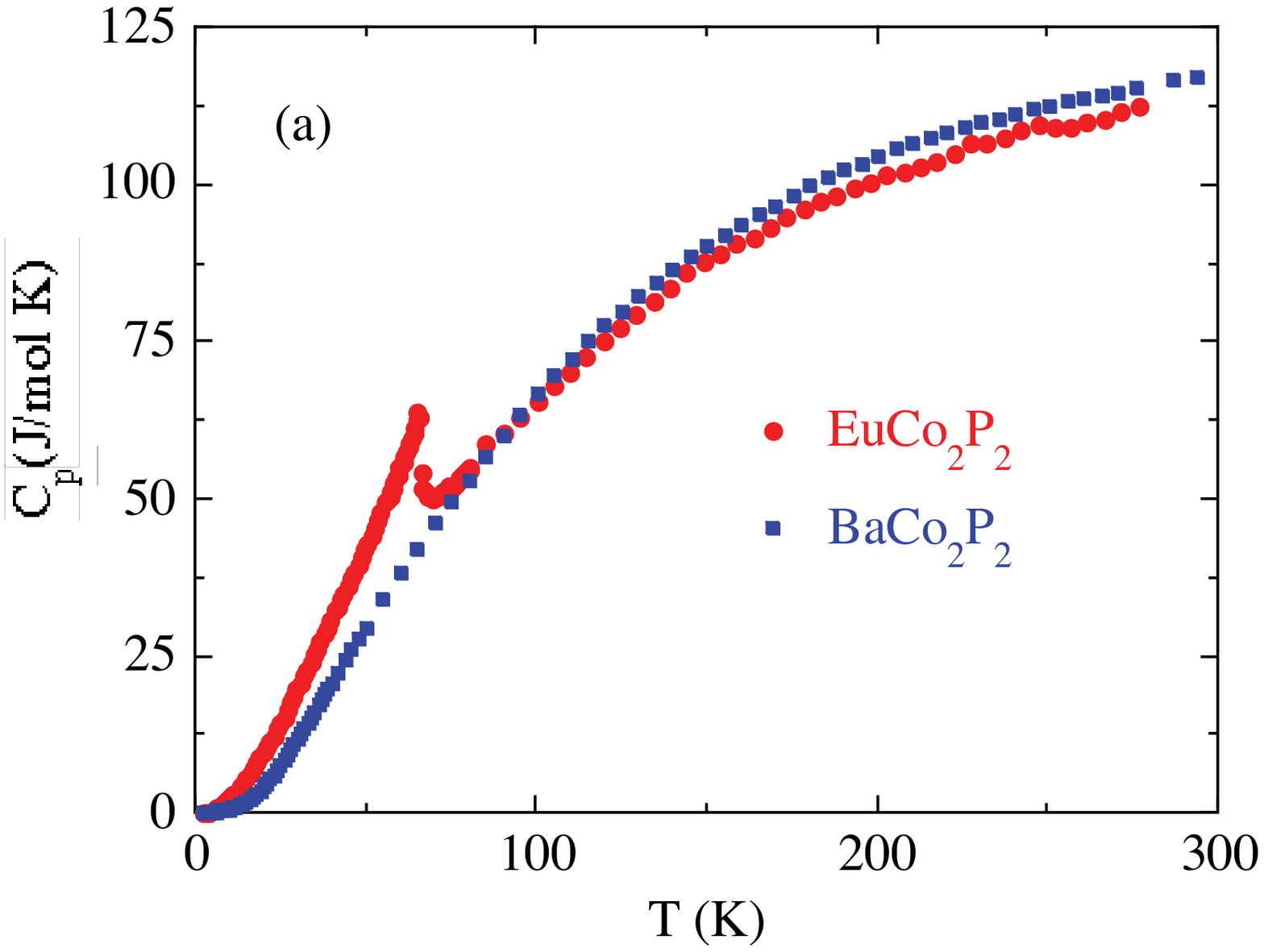}
\includegraphics[width=2.85in]{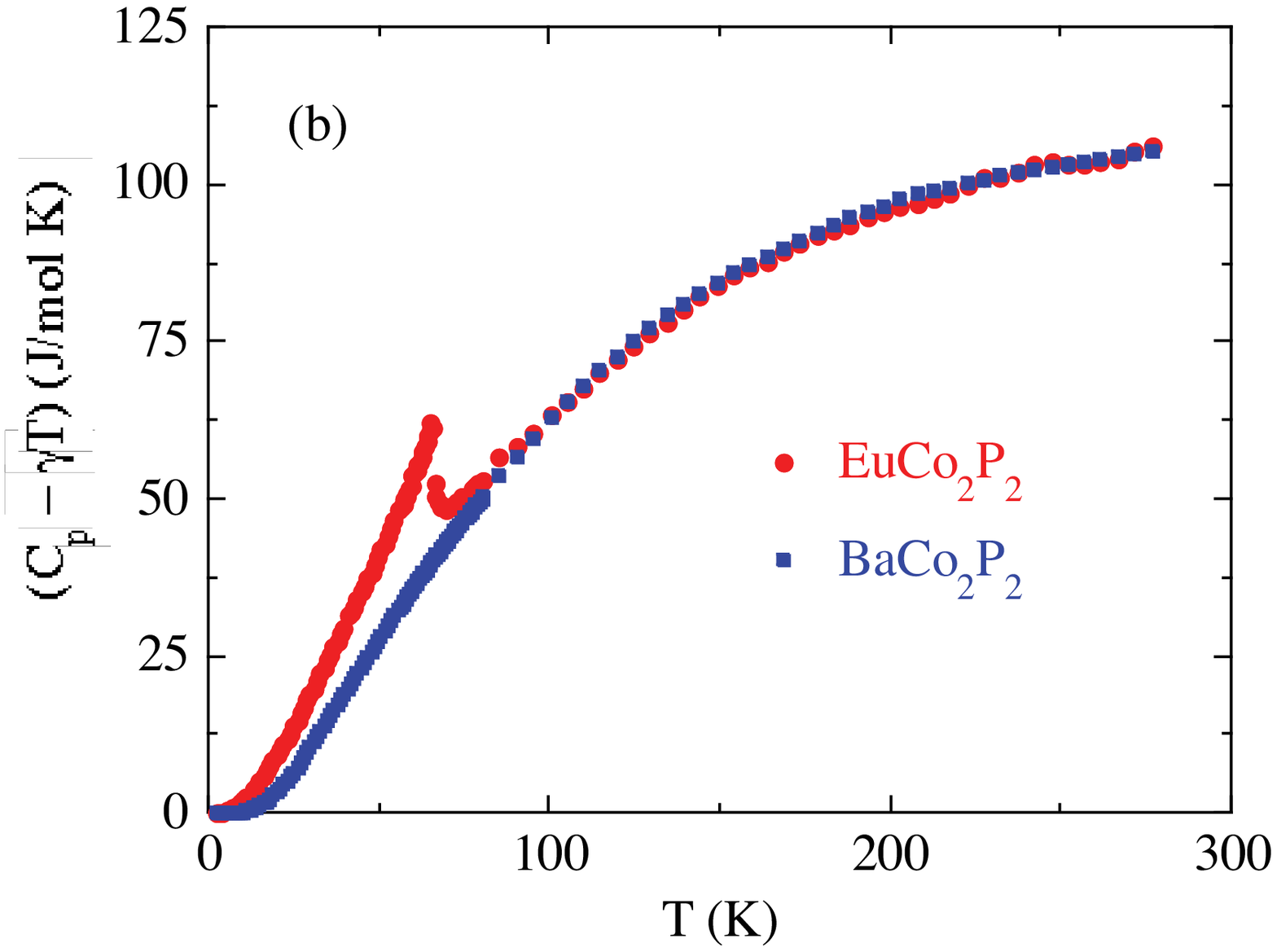}
\caption {(Color online) (a)  $C_{\rm p}$ versus $T$ for \ecp\ and ${\rm BaCo_2P_2}$.  (b)~The difference $C_{\rm p}-\gamma T$ versus $T$ for \ecp\ and ${\rm BaCo_2P_2}$, where $\gamma$ is the Sommerfeld electronic heat capacity coefficient for the respective compound.}
\label{Fig:CpEuBascaled}
\end{figure}

\begin{figure}
\includegraphics[width=2.85in]{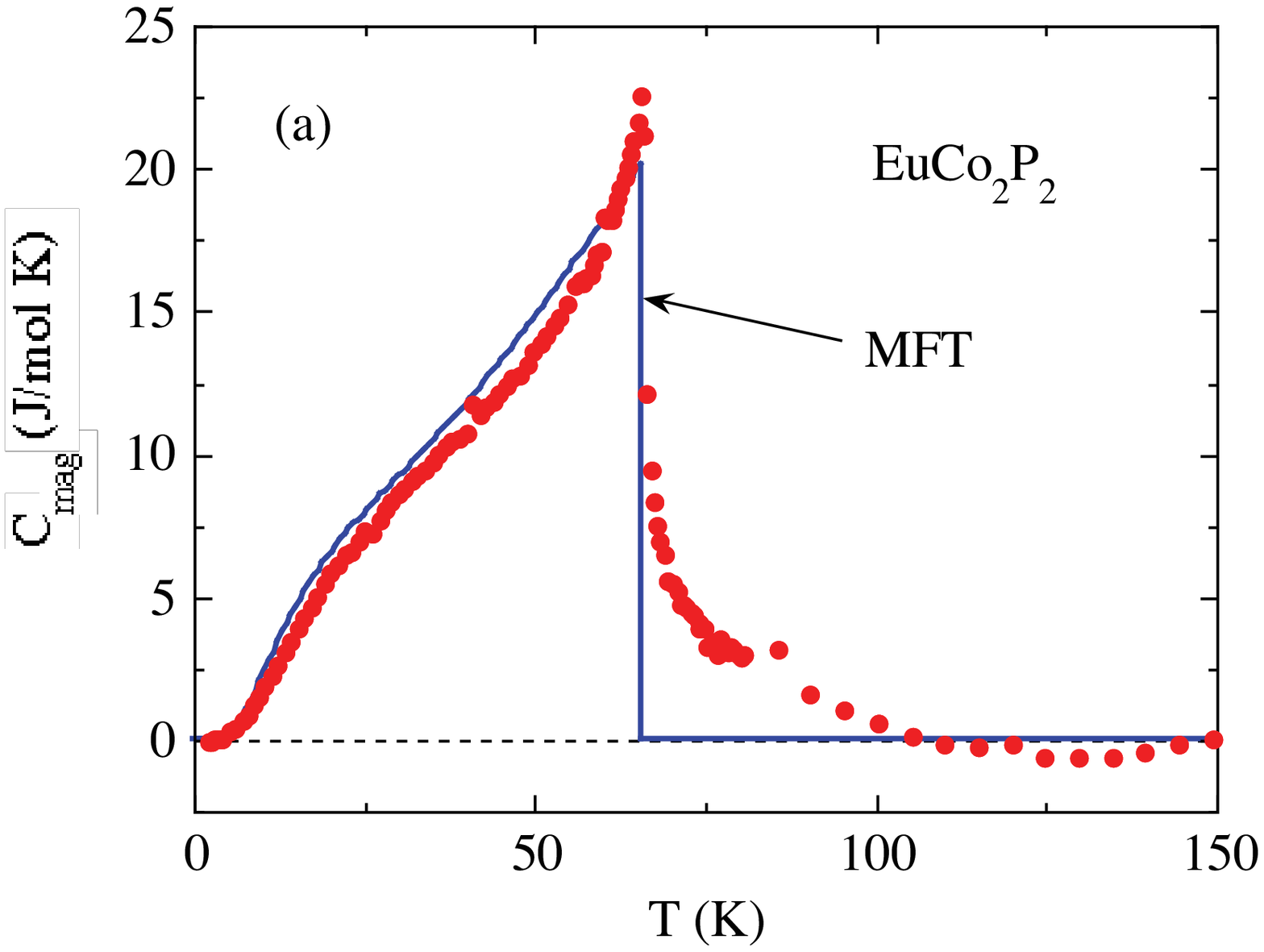}
\includegraphics[width=2.85in]{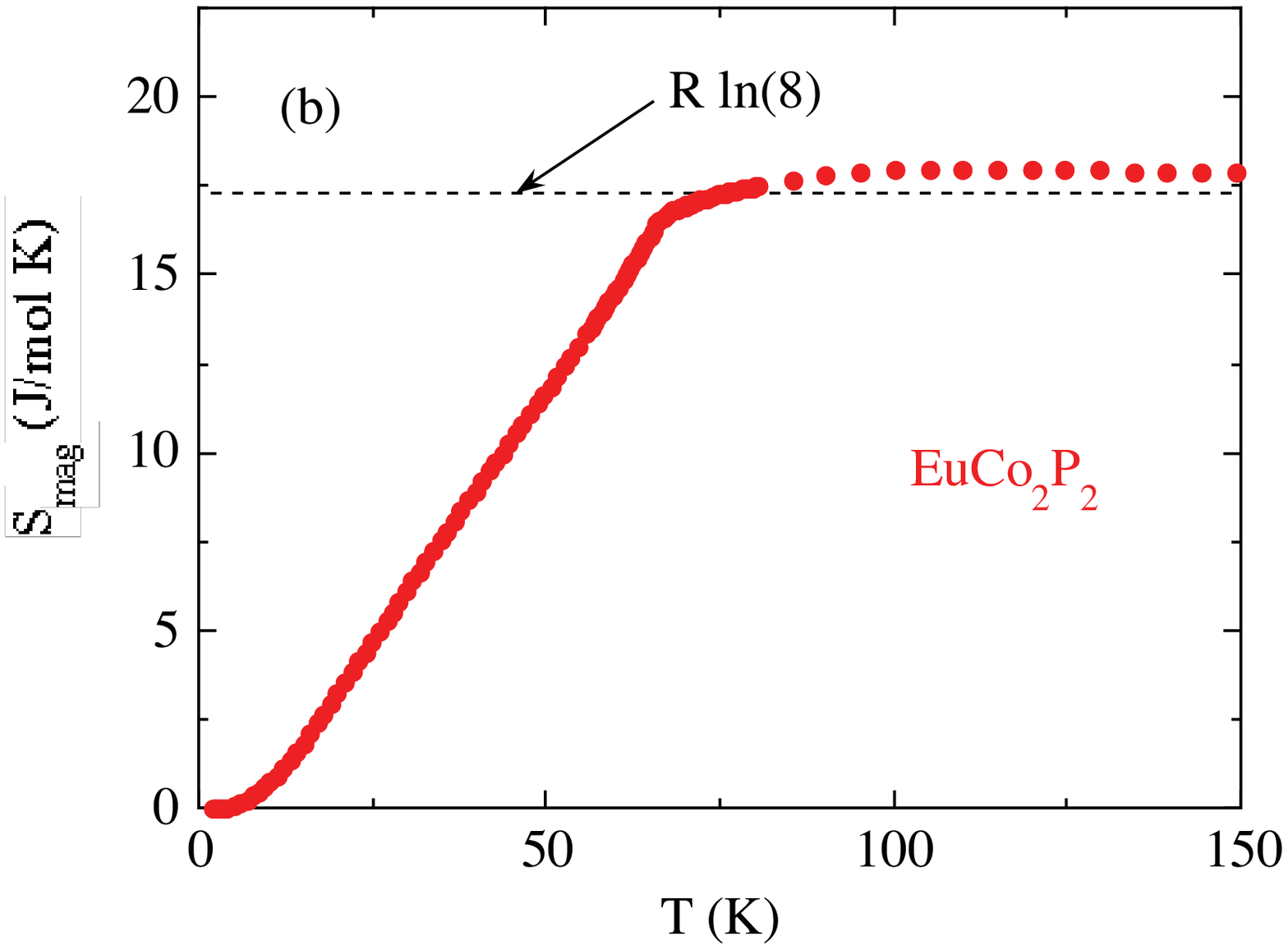}
\caption {(Color online) (a) $C_{\rm mag}$  versus $T$ for \ecp\ (open circles).  The solid curve is the prediction of MFT\@.  (b)~$S_{\rm mag}$  versus $T$ for \ecp.  The horizontal dashed line is the theoretical high-$T$ limit $S_{\rm mag} = R\ln(2S+1)$ for $S=7/2$.}
\label{Fig:Cmag}
\end{figure}

Our crystals of \ecp\ were grown in Sn flux as described previously \cite{Reehuis1992}, whereas a polycrystalline sample of \bcp\ was prepared by solid-state reaction.  Magnetization~$M$ data were obtained using the SQUID magnetometer in a Quantum Design, Inc., MPMS system for applied magnetic fields $H \leq 5.5$~T and using a vibrating sample magnetometer in a PPMS system for high-field $M(H)$ isotherm measurements up to 14~T\@. $C_{\rm p}(T)$ and $\rho(T)$ data were also obtained using the PPMS system.

\begin{figure}
\includegraphics[width=2.85in]{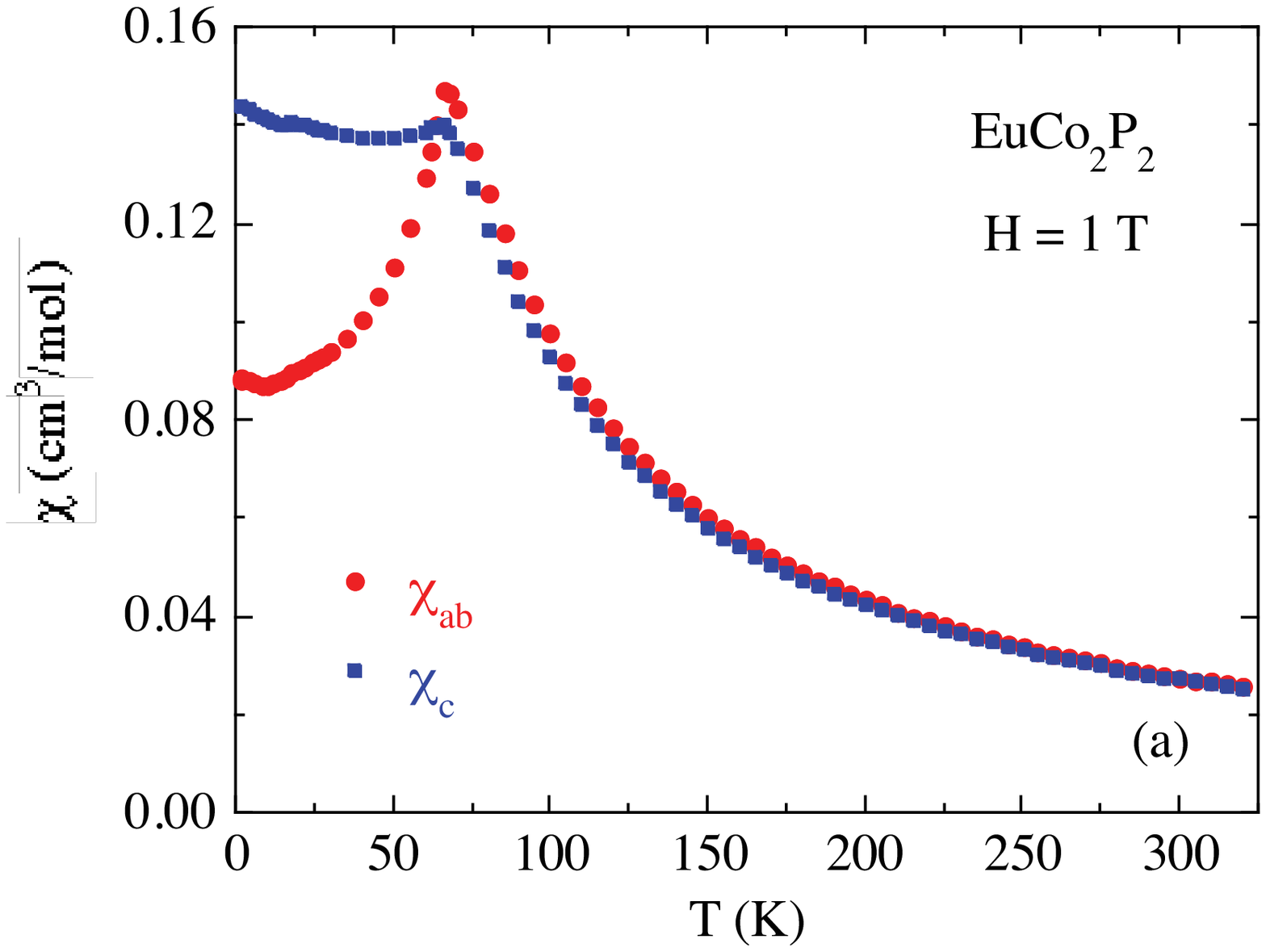}
\includegraphics[width=2.85in]{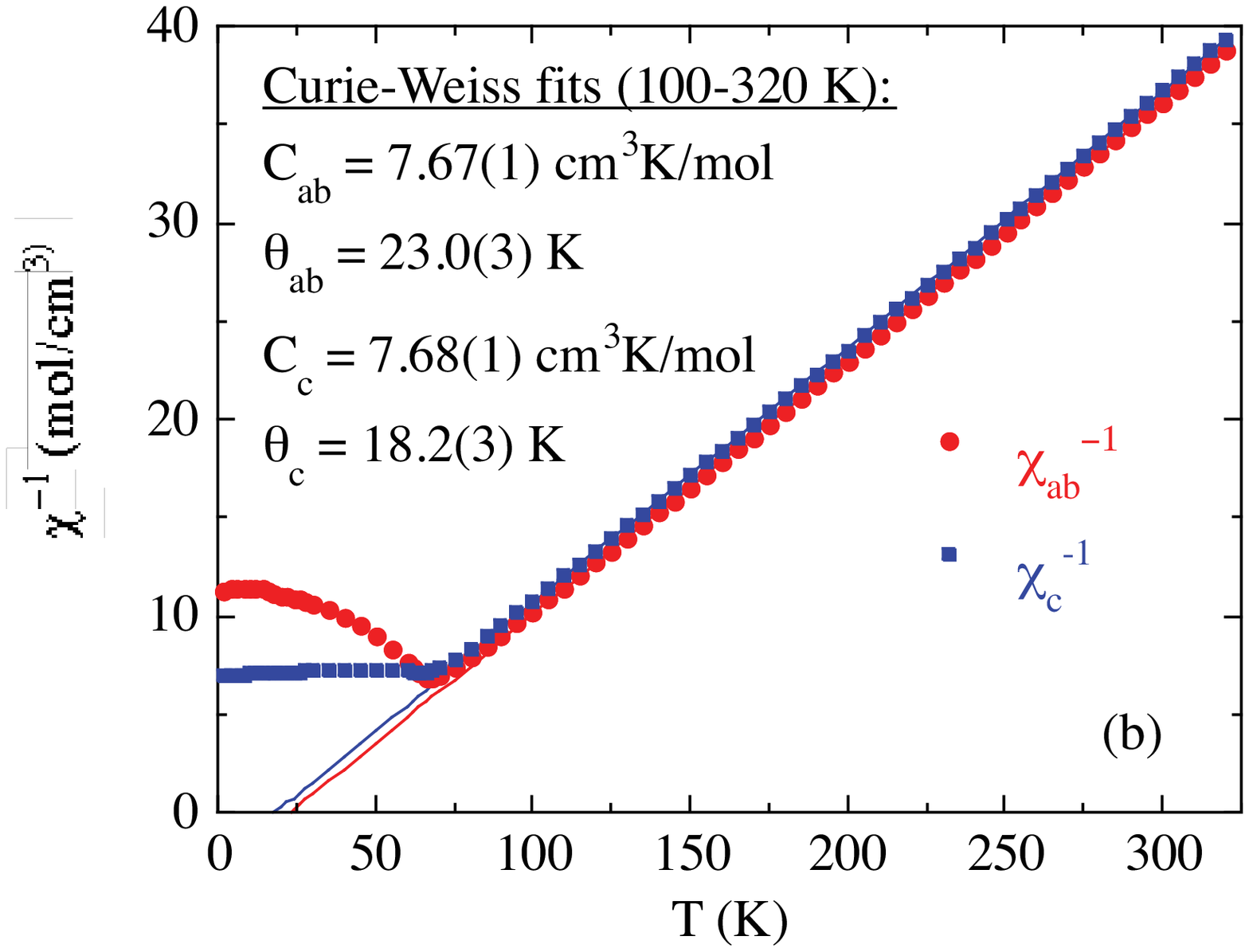}
\caption {(Color online) (a) $\chi\equiv M/H$ versus $T$ for single-crystal \ecp\ for $H\parallel c$~axis and $H\parallel ab$~plane with $H=1$~T\@.  (b)~$\chi^{-1}$ versus~$T$ for both $H\parallel c$ and $H\parallel ab$.  The data above 100~K are fitted by the Curie-Weiss law (straight lines) with the fit parameters listed.}
\label{Fig:chi_vs_T}
\end{figure}

$C_{\rm p}$ data for \ecp\ and \bcp\ are plotted versus~$T$ in Fig.~\ref{Fig:CpEuBascaled}(a).  The data for \bcp\ are used to correct for $C_{\rm p}$ contributions in \ecp\ from other than the Eu magnetism.  A pronounced peak is seen for \ecp\ at $T_{\rm N} = 65.7(1)$~K that is comparable with the previously-reported $T_{\rm N}$ values \cite{Morsen1988, Reehuis1992, Nakama2010}.  At low~$T$ the data for \ecp\ and \bcp\ follow $C_{\rm p} = \gamma T + \beta T^3$, where $\gamma({\rm EuCo_2P_2}) = 23.7(5)$ and $\gamma({\rm BaCo_2P_2}) = 37.3(3)~{\rm mJ/(mol~K^2)}$ (not shown).  The nonzero values of the Sommerfeld coefficients $\gamma$ indicate metallic character for both compounds.  Shown in Fig.~\ref{Fig:CpEuBascaled}(b) are the $C_{\rm p}$ data corrected for the respective electronic $\gamma T$ terms.  One sees that the $C_{\rm p} - \gamma T$ data for the two compounds are now nearly identical above $\approx100$~K\@.  To eliminate the residual average deviation of the data between the two compounds in the 100--300~K temperature range we multiplied the $C_{\rm p} - \gamma T$ data for \ecp\ by 1.0046.  Then taking the difference between the resulting data for \ecp\ and the $C_{\rm p} - \gamma T$ data for \bcp\ yields the magnetic heat capacity $C_{\rm mag}(T)$ for \ecp\ in Fig.~\ref{Fig:Cmag}(a).  The nonzero $C_{\rm mag}$ for $T_{\rm N} < T\lesssim 100$~K is due to weak dynamic short-range magnetic ordering of the Eu spins above $T_{\rm N}$\@.  The MFT prediction \cite{Johnston2015} for $C_{\rm mag}(T)$ for $S=7/2$  and $T_{\rm N} = 65.7$~K is shown as the solid blue curve in Fig.~\ref{Fig:Cmag}(a).  The agreement is quite good, apart from the 30~K region above $T_{\rm N}$ where the short-range magnetic ordering there is not included in the MFT prediction.

The magnetic entropy $S_{\rm mag}(T)$ is calculated using $S_{\rm mag}(T) = \int_0^T [C_{\rm mag}(T)/T]dT$ and the result is shown in Fig.~\ref{Fig:Cmag}(b).  The high-$T$ limit for a mole of spins $S=7/2$ is $R\ln(2S+1)$, where $R$ is the molar gas constant, as shown by the horizontal dashed line in Fig.~\ref{Fig:Cmag}(b).  One sees that the high-$T$ $S_{\rm mag}(T)$ data for \ecp\ closely approach this value.  The small residual deviation is likely due to a small inaccuracy in the background $C_{\rm p}$ subtraction.  The short-range magnetic ordering seen in $C_{\rm mag}$ at $T>T_{\rm N}$ in Fig.~\ref{Fig:Cmag}(a) represents only a small fraction of the total entropy of the disordered spin system, since the change in $S_{\rm mag}$ from $T_{\rm N}$ to 100~K is found from Fig.~\ref{Fig:Cmag}(b) to be only about 7\% of the disordered entropy $R\ln(8)$.

We also carried out in-plane $\rho(T)$ measurements on a \ecp\ crystal and the results (not shown) are similar to those in \cite{Nakama2010}.  In particular, on cooling we find that the $\rho(T)$ shows a sharp increase in slope at $T_{\rm N} =  66(1)$\,K, consistent with our $C_{\rm mag}$ measurement of $T_{\rm N}$ above.

The $\chi\equiv M/H$ data are presented versus~$T$ for $H = 1$~T in Fig.~\ref{Fig:chi_vs_T}(a) for both $H\parallel c$~axis and $H\parallel ab$~plane.  The discontinuities in the slope versus $T$ occur at $T_{\rm N} \approx 66.5$~K, in agreement with the previous reports \cite{Morsen1988, Nakama2010, Reehuis1992}.  The $\chi(T)$ data for $H\parallel c$ become nearly independent of $T$ below $T_{\rm N}$, consistent with the MFT prediction \cite{Johnston2015} for the perpendicular susceptibility of the helical AFM structure of \ecp\ in which the ordered moments are oriented within the $ab$~plane \cite{Reehuis1992} .  The data for $H\parallel ab$ have a nonzero limit at $T\to0$, again consistent with the planar noncollinear AFM structure \cite{Johnston2012, Johnston2015, Anand2015}.  

The inverse susceptibility $\chi^{-1}$ versus~$T$ is shown in Fig.~\ref{Fig:chi_vs_T}(b).  The data from 100~K to 320~K were fitted by the Curie-Weiss law, yielding the Curie constants and Weiss temperatures listed in the figure.  Since the Curie constants are the same for the two field directions, the effective moments derived from them are also the same, $\mu_{\rm eff} = 7.84~\mu_{\rm B}$/f.u., which agrees within about 1\% with the value $7.94\,\mu_{\rm B}$/f.u.\ expected for Eu$^{+2}$ with $S = 7/2$ and $g=2$.  The difference $\theta_{ab}-\theta_c$ arises mainly from magnetic dipole interactions between the Eu spins \cite{Johnston2016}.  The spherical average of the fitted Weiss temperatures is $\theta_{\rm p,ave} = 21.4(3)$~K, in agreement with the previous result $\theta_{\rm p} = 22(2)$~K \cite{Morsen1988} for a polycrystalline sample.

\begin{figure}
\includegraphics[width=2.85in]{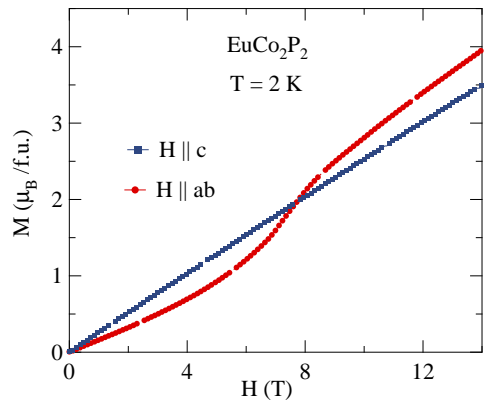}
\caption {(Color online) $M$ versus~$H$ at $T=2$~K up to high fields for $H\parallel c$ and $H\parallel ab$.}
\label{Fig:MH_2K}
\end{figure}

High-field $M(H)$ isotherm data at $T=2$~K for $H\parallel c$~axis and $H\parallel ab$~plane are shown in Fig.~\ref{Fig:MH_2K}.  The data for $H\parallel c$ are nearly linear in field, in agreement with the high-field behavior expected within MFT for a field perpendicular to the ordering plane of a noncollinear AFM \cite{Johnston2015}.  Extrapolating the $M_c(H)$ data in Fig.~\ref{Fig:MH_2K} to the saturation moment $\mu_{\rm sat}=gS\mu_{\rm B} = 7~\mu_{\rm B}$/Eu gives the critical field $H_{\rm c}(T\to0) \sim 28$~T at which a second-order transition to the paramagnetic state occurs with increasing $H$ \cite{Johnston2015}.  For the in-plane $H\parallel ab$, the $M_{ab}(H)$ data in Fig.~\ref{Fig:MH_2K} show a metamagnetic transition at $\approx 8$~T\@.  The significant width of this transition is consistent with a noncollinear AFM structure in $H=0$ with the ordered moments aligned in the $ab$~plane.  The $M_{ab}(H)$ behavior at high~$H$ has not yet been worked out in detail for a helical AFM\@.

\begin{figure}
\includegraphics[width=2.85in]{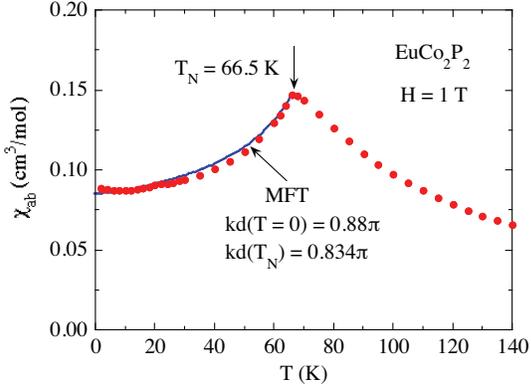}
\caption {(Color online) $\chi_{ab}(T)$ versus $T$ for $H\parallel ab$~plane in $H=1$~T\@.  The fit by MFT for $\chi_{ab}(T)$ of a helix in Eqs.~(\ref{Eqs:Chixy}) for $T\leq T_{\rm N}$ is shown as the solid blue curve.}
\label{Fig:Chi_1T_EuCo2P2_TN66_ka88}
\end{figure}

An expanded plot of $\chi_{ab}(T)$ for \ecp\ from Fig.~\ref{Fig:chi_vs_T}(a) is shown in Fig.~\ref{Fig:Chi_1T_EuCo2P2_TN66_ka88}.   The normalized in-plane susceptibility $\chi_{ab}(T \leq T_{\rm N})/\chi(T_{\rm N})$ for a planar noncollinear helical AFM system is given within our formulation of MFT by \cite{Johnston2012,Johnston2015}
\bse
\label{Eqs:Chixy}
\begin{equation}
\frac{\chi_{ab}(T \leq T_{\rm N})}{\chi(T_{\rm N})}=  \frac{(1+\tau^*+2f+4B^*)(1-f)/2}{(\tau^*+B^*)(1+B^*)-(f+B^*)^2},
\label{eq:Chi_plane}
\end{equation}
where
\begin{equation}
B^*=  2(1-f)\cos(kd)\,[1+\cos(kd)] - f,
\label{eq:Bstar}
\end{equation}

\be
t =\frac{T}{T_{\rm N}},\quad f=\frac{\theta_{\rm p}}{T_{\rm N}}, \quad\tau^*(t) = \frac{(S+1)t}{3B'_S(y_0)}, \quad y_0 = \frac{3\bar{\mu}_0}{(S+1)t}, 
\ee
\ese
the ordered moment versus $T$ in $H=0$ is denoted by $\mu_0$, the reduced ordered moment $\bar{\mu}_0 = \mu_0/\mu_{\rm sat}$ is determined by solving
$\bar{\mu}_0 = B_S(y_0)$, $B'_S(y_0) = [dB_S(y)/dy]|_{y=y_0}$ and our unconventional definition of the Brillouin function $B_S(y)$ is given in \cite{Johnston2012, Johnston2015}.

We fitted the data in Fig.~\ref{Fig:Chi_1T_EuCo2P2_TN66_ka88} for $T\leq T_{\rm N}$ by Eqs.~(\ref{Eqs:Chixy}) using $S=7/2$, $T_{\rm N} = 66.5$~K, $\chi(T_{\rm N}) = 0.147\,{\rm cm^3/mol}$ and  $f = \theta_{ab}/T_{\rm N} = 0.346$ where $\theta_{ab} = 23.0$~K\@.   For $kd(T)$ we used the value $kd(T=64~{\rm K}) = 0.834\pi$~rad from Eq.~(\ref{Eq:kdNeuts}) \cite{Reehuis1992}.  In order to fit the lowest-$T$ data in Fig.~\ref{Fig:Chi_1T_EuCo2P2_TN66_ka88}, we used $kd(T=0) = 0.88\pi$~rad, which is comparable to the value at 15~K in Eq.~(\ref{Eq:kdNeuts}) \cite{Reehuis1992}.  For intermediate temperatures $kd$ was linearly interpolated between these two values.  The $\chi_{ab}(T\leq T_{\rm N})$ thus obtained from MFT is plotted versus~$T$ for $T\leq T_{\rm N}$ as the solid curve in Fig.~\ref{Fig:Chi_1T_EuCo2P2_TN66_ka88}.  The fit is seen to be in good agreement with the data.  

We now estimate the intralayer and interlayer Heisenberg exchange interactions within a minimal $J_0$-$J_{z1}$-$J_{z2}$ MFT model for a helix \cite{Johnston2012, Johnston2015}, where $J_0$ is the sum of all Heisenberg exchange interactions of a representative spin to all other spins in the same spin layer perpendicular to the helix ($c$) axis, $J_{z1}$ is the sum of all interactions of the spin with spins in an adjacent layer along the helix axis, and $J_{z2}$ is the sum of all interactions of the spin with spins in a second-nearest layer \cite{Johnston2012, Johnston2015}.  Within this model $kd$, $T_{\rm N}$ and $\theta_{{\rm p}}$ are related to these exchange interactions by \cite{Johnston2012, Johnston2015}
\bse
\label{Eqs:J0J1zJ2z}
\bea
&&\cos(kd) = -\frac{J_{z1}}{4J_{z2}},\label{Eq:coskd}\\*
T_{\rm N} &=& -\frac{S(S+1)}{3k_{\rm B}} \big[J_0 + 2J_{z1}\cos(kd)\nonumber\\*
&& \hspace{0.9in} +\ 2J_{z2}\cos(2kd)\big], \label{eq:TN}\\*
\theta_{\rm p} &=& -\frac{S(S+1)}{3k_{\rm B}} \left(J_0+2J_{z1}+2J_{z2}\right)\label{eq:thetap},
\eea
\ese
where a positive (negative) $J$ corresponds to an AFM (FM) interaction.  Using $S = 7/2,\ T_{\rm N} = 66.5\ {\rm K},\ \theta_{\rm p}=23.0$~K, the average $kd=0.857\pi$~rad of the two $kd$ values in Fig.~\ref{Fig:Chi_1T_EuCo2P2_TN66_ka88} and solving Eqs.~(\ref{Eqs:J0J1zJ2z}) for the three exchange constants, one obtains
\bea
J_0/k_{\rm B} &=& -9.7~{\rm K},\quad J_{z1}/k_{\rm B} = 2.1~{\rm K},\\*
&&J_{z2}/k_{\rm B} = 0.6~{\rm K},\nonumber
\eea
where the above variation in $kd$ with $T$ for $0<T<T_{\rm N}$ is found to have a minimal effect on the derived $J$'s.  As anticipated above, from the FM-like $\theta_{\rm p}$ the net exchange constant $J_0+2J_{z1}+2J_{z2} = -4.4$~K is FM, and the out-of-plane exchange constants $J_{z1}$ and $J_{z2}$ are both AFM.

In summary, we demonstrated from measurements on single crystals that \ecp\ is a textbook example of a noncollinear MFT helical AFM where its $\chi$ at $T\leq T_{\rm N}$ is well fitted by the predictions of MFT\@.  The $C_{\rm mag}(T)$ data also agree well with the MFT prediction apart from weak dynamic short-range AFM ordering above $T_{\rm N}$ that is not taken into account by MFT\@.  From the values of $S$, $kd$, $T_{\rm N}$ and $\theta_{\rm p}$ we extracted values of the Heisenberg exchange constants within the $J_0$-$J_{z1}$-$J_{z2}$ MFT model.  It would be interesting to compare these exchange constant values with the predictions of electronic structure calculations for this model system.


This research was supported by the U.S.~Department of Energy, Office of Basic Energy Sciences, Division of Materials Sciences and Engineering.  Ames Laboratory is operated for the U.S.~Department of Energy by Iowa State University under Contract No.~DE-AC02-07CH11358.


\end{document}